\documentclass[11pt]{article}
\usepackage{graphicx}
\usepackage{amssymb}
\usepackage{graphics}

\input{tcilatex}

\begin{document}

\title{Comment to \textquotedblleft Observation of the neutron radiative
decay\textquotedblright\ by R.U. Khafizov et al., published in JETP Letters
83 (2006) 5 (Pis'ma v ZhETF 83 (2006) 7)}
\author{N. Severijns$^{1}$, O. Zimmer$^{2}$, H.-F. Wirth$^{2}$ and D. Rich$%
^{2}$  \\
$^{1}$Katholieke Universiteit Leuven, B-3001 Leuven, Belgium\\
$^{2}$Technische Universit\"{a}t M\"{u}nchen, 85747 Garching, Germany}
\maketitle

Herewith we formally deny any responsibility for the content of the
commented paper. The manuscript was submitted for publication without
informing at least four of the other authors, viz. N. Severijns, O. Zimmer,
H.-F. Wirth and D. Rich. This violates our rights as collaborators. The
analysis presented and the manuscript itself have not been discussed and
have also not been approved by the entire collaboration prior to submission.
Besides this formal incorrectness, we also criticise the content of the
paper (low quality and premature). Not only the interpretation, but already
the presentation of data is not comprehensive and does not fulfill minimum
scientific standards. These views are shared by J. Byrne who is
collaborating on the project as well, but was not mentioned on the paper.

The coincidence spectrum shown in fig. 5 contains a forest of peaks. Only
the two major peaks are explained. The highest peak is called
\textquotedblleft zero or false coincidences\textquotedblright\ with
reference to a publication by another group, where this peak is not
discussed at all. To our understanding it might be due to a physics event as
for example detection by the proton detector of an electron bremsstrahlung
photon created in the electron detector, coincident with detection of an
electron by the electron detector (as seems to be suggested in the paper),
or cross-talk of electronics of the corresponding detectors, or some other
reason. No attempt for explanation is made for the smaller peaks in Fig. 5.

Figure 6 seems to show a spectrum of triple coincidences between three
different detectors. Without that it is mentioned explicitly in the figure
caption, the explanations given seem to indicate that the horizontal axis
represents the time between an event in the electron detector and an event
in the proton detector. It was said in the text that the feature in
\textquotedblleft channel 120 in Fig. 6\textquotedblright\ contains
coincidences between decay electrons and protons, along with an event in one
of the gamma detectors. However, no width of the coincidence window is
stated. Further, since exactly this type of events was announced to be used
for determination of the branching ratio of radiative neutron decay, the
statement that the analysis of the branching ratio was based on the leftmost
peak with maximum in channel 102 remains completely incomprehensible. Also,
no attempt for explanation is made for the broad bump with maximum at about
channel 165 in Fig. 6.

Besides these deficiencies of the presented treatment it is obvious that the
detector setup allows for many types of events which were not discussed.
Without a careful analysis of backgrounds one just cannot extract any
quantitative information about neutron decay, and in particular not
seriously extract a branching ratio of a hitherto unuobserved effect. We
therefore do not endorse the claim of the first author to have observed
radiative neutron decay. 

\end{document}